\documentclass[twoside]{article}
\usepackage{fleqn,espcrc2}
\usepackage{psfig}
\title{The antikaon potential in nuclear matter at 
finite momentum\thanks{Supported 
by Forschungszentrum J\"ulich}}
\author{A. Sibirtsev and W. Cassing \\ \vspace{5mm}
Institut f\"ur Theoretische Physik, Universit\"at Giessen \\
D-35392 Giessen, Germany}

\begin{document}

\begin{abstract}
We study the antikaon potential at finite momentum for 
proton and neutron nuclear matter. Our approach is based on 
the momentum dependence of the real part of the total $K^-N$
scattering amplitude, which is fixed in free space by experimental data, 
using the dispersion relation approach. We decompose
the contributions to the scattering amplitude from different 
processes and consider the $\Lambda$(1405) and 
$\Sigma$(1385) to be dissolved in nuclear 
matter at density $\rho_0$. Within this approach the 
$K^-$ potential is found to be attractive
up to high kaon momenta. Our result is in line with the data on the
antikaon potential evaluated from the analysis of 
kaonic atoms (low momenta), heavy-ion 
(interdmediate momenta) and proton-nucleus (high momenta) collisions.
\vspace{1pc}
\end{abstract}
\maketitle

The antikaon potential in nuclear matter at finite relative momentum at
present is a question of vivid interest, which is partly discussed in a
controversal manner. The analysis~\cite{Gal} of data on kaonic 
atoms leads to an antikaon potential of $\simeq$--180~MeV at normal 
nuclear density $\rho_0$,
while the studies~\cite{IonsT} of $K^-$-meson production from heavy-ion
collisions~\cite{IonsE} suggest an attractive potential  
$\simeq$--100$\div$120~MeV.
We have proposed in Ref.~\cite{Sibirtsev} to attribute 
this discrepancy to the
momentum dependence of the antikaon potential, since the kaonic
atoms explore stopped antikaons with $p_K{\approx}$0, while the heavy-ion
experiments have probed the range 300${\le}p_K{\le}$600~MeV/c. The 
evaluation of the momentum-dependent potential is a rigorous problem and 
substantially depends on the model applied. Here we
discuss a dispersive approach in  which the uncertainties are under 
theoretical and experimental control.  

Within the low density theorem the real part of the antikaon potential 
at baryon density $\rho_B$ is given in terms of the real forward
$K^-$-nucleon scattering amplitude $D$ as
\begin{equation}
U(\rho_B, E_K) = - \frac{2\pi}{m_K} \rho_B \ D(E_K),
\label{low}
\end{equation}
where $E_K$ is the kaon energy relative to the nuclear matter
rest frame. It is important to note that $D$ here is the total scattering
amplitude including all possible processes available at 
given antikaon energy
$E_K$. It is obvious that with increasing $E_K$ the number of open 
reaction channels becomes very large such that all channels cannot 
be calculated separately anymore. 
The contribution from the individual reaction
channels then can be controlled only by the relative 
saturation of the total 
cross section. In case of the $K^-p$ interaction the 
$\Lambda(1405)$ resonance dominates at low energy,
but apart of its specific role there are contributions from other 
channels as indicated by the data on the total 
cross section~\cite{PDG}. Thus models relying exclusively 
on the $\Lambda(1405)$ properties
in the medium have to be taken with great care.

An alternative way is to evaluate the total real forward 
scattering amplitude $D$ from the total cross section directly, which 
is controlled by experimental data. 
The advantages of this method are i) an absolute completeness
of the scattering amplitude $D$ with respect to all available 
channels, and 
ii) an independence on the hyperon resonance properties in the medium at
sufficiently high density where the latter are 'melted'.
In free space the real part of the forward total
scattering amplitude $D$ can be evaluated from the dispersion 
relation as~\cite{Theory}

\begin{eqnarray}
D(E_K)=D(E_K{=}0)+E_K\sum\limits_{j=1}^2 I_j(E_K) \nonumber \\
+ \sum\limits_{Y=\Lambda}^{\Sigma}
\frac{g^2_{KNY}}{16\pi\  m_N^2 } 
\frac{E_K \ [m_K^2-(m_Y -m_N)^2]}{E_Y(E_K-E_Y)}
\label{disp}
\end{eqnarray}
where $E_\Lambda$=64~MeV, $E_\Sigma$=155~MeV while $I_1$ stands
for the contribution from the unphysical domain $E_K{<}m_K$:
\begin{equation}
I_1{=}\frac{1}{\pi m_N}\!\!\!
\int\limits_{E_{\Lambda \pi}}^{\ \  m_K}
\frac{d\omega A (\omega)\sqrt{m^2_N+m^2_K+2m_N\omega}}
{\omega (\omega - E_K)}
\end{equation}
with $E_{\Lambda \pi}$=239~MeV. $A(\omega)$ is the imaginary 
part of the forward $K^-N$ scattering amplitude that stems explicitly from
the $\Lambda{(1405)}$ and $\Sigma{(1385)}$ resonances. The
contribution from the physical region is given as
\begin{equation}
I_2{=}\frac{1}{4\pi^2}\!\!\int\limits_{m_K}^{\infty}
\!\!\frac{d\omega \sqrt{\omega^2-m_K^2}}{\omega}
\left[ \frac{
\sigma^-(\omega )}{\omega -E_K}{-}\frac{\sigma^+(\omega )}
{\omega{+}E_K}\right]
\end{equation}
where $\sigma^-$ and $\sigma^+$ denote the total $K^-N$ and $K^+N$
cross sections, respectively, which are fixed by experimental
data~\cite{PDG}.

\begin{figure}[h]
\vspace{-5mm}\hspace{-2mm}
\psfig{file=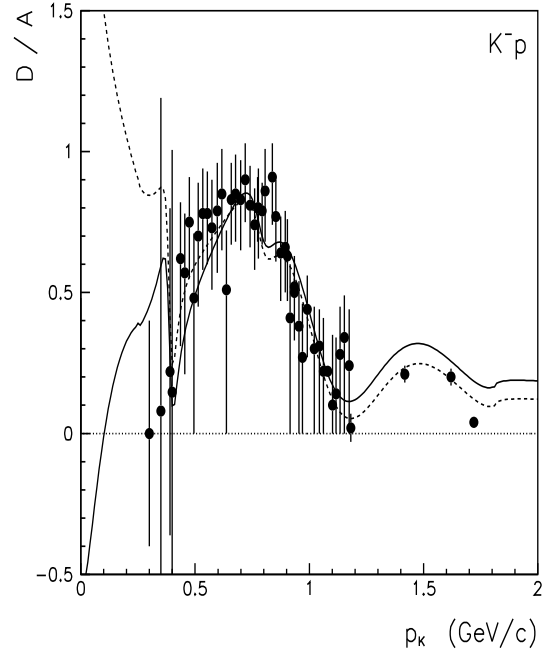,width=8.1cm,height=10cm}
\vspace{-15mm}
\caption{The ratio of the real $D$ to imaginary part $A$ 
of the forward  $K^-p$ scattering amplitude in comparison 
to the data from~\protect\cite{Data}. The lines show the dispersion
calculations: full (solid) and without the contribution $I_1$ from 
the unphysical domain (dashed), but adjusted to data (dotted) 
by D($E_K$=0).}
\label{regen2}
\vspace{-5mm}
\end{figure}

The dispersion relation can be evaluated by fitting 
$D(0)$, $g_{KN\Lambda}$ and $g_{KN\Sigma}$ to the data on the real
$K^-N$ and $K^+N$ forward scattering amplitudes while taking
$A(\omega)$ from the $M$-matrix low energy solution~\cite{Matrix}. 
We note that the 
relation between the $K^-N$ and $K^+N$ scattering amplitudes
is given by crossing symmetry. 

The solid line in Fig.~\ref{regen2} shows the ratio of the 
real $D(p_K)$  to imaginary $A(p_K)$ part of the forward 
scattering amplitudefrom~\cite{Sibirtsev} together with the available
$K^-p$ experimental results. To illustrate the sensitivity to the
contribution from the unphysical domain we also show (dashed 
line) the result when neglecting $I_1$ and varying only
$D(0)$. In principle, the real part of the $K^-p$ scattering 
amplitude is quite sensitive to $A(\omega)$, but in terms of 
the available data (apart from the two lowest energy points) it 
cannot be fixed uniquely. 

\begin{figure}[h]
\vspace{-6mm}\hspace{-2mm}
\psfig{file=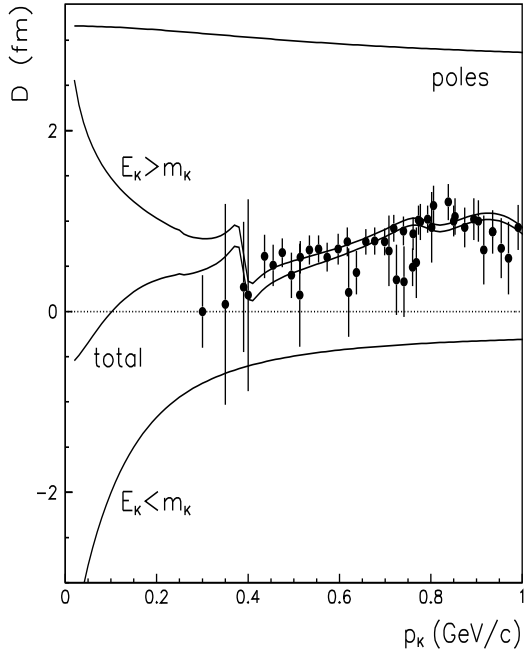,width=8.1cm,height=10cm}
\vspace{-15mm}
\caption{The contribution to the real 
forward $K^-p$ scattering amplitude from different terms
in Eq.~\ref{disp}.}
\label{regen2b}
\vspace{-6mm}
\end{figure}

\begin{figure}[h]
\vspace{-9mm}\hspace{-5mm}
\psfig{file=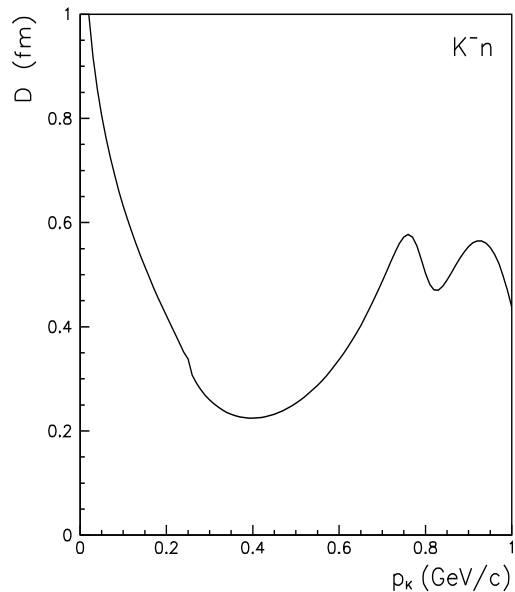,width=8.2cm,height=9cm}
\vspace{-10mm}
\caption{The real part of the forward $K^-n$ scattering amplitude
as a function of the antikaon momentum $p_K$.}
\label{regen2a}
\vspace{-2mm}
\end{figure}

Furthermore, it is obvious that the amplitude $D$ becomes 
negative for $p_K{<}$100~MeV/c due to the contribution 
(for $E_K{<}m_K$) from the $\Lambda$(1405) resonance. Fig.~\ref{regen2b} 
shows the contribution to the total forward $K^-p$ scattering 
amplitude from the pole term, the unphysical ($I_1$) as 
well as from the physical region ($I_2$). Note that
apart from the $\Lambda$(1405) contribution and the 
boundary $D(0)$=--2.73~fm
the other contributions to the real forward scattering amplitude 
are positive.
Furthermore, the $\Lambda$(1405) resonance does not couple to the $K^-n$
system and as shown in Fig.~\ref{regen2a} the real
part of the $K^-$-neutron scattering amplitude remains positive, i.e. the
$K^-$ potential in neutron matter is  attractive.

Now, it is by far not obvious that the forward scattering amplitude
in nuclear matter is the same as in free space, but it is clear 
that all terms contributing to $D$ in the vacuum should be 
also considered in calculations for nuclear matter. The experimental
result on kaonic atoms~\cite{Gal} indicates that the $K^-N$ forward
scattering amplitude is positive at $p_K{\approx}0$ already at the rather
low nuclear density which the antikaon experiences before decay.
This implies that the amplitude $D$ is modified in matter already 
at rather low densities, which was attributed in Ref.~\cite{Brown} 
to the 'melting' of 
the  $\Lambda$(1405) since this resonance gives repulsion. 
In fact, dynamical calculations~\cite{Dynamic} 
on the low energy $K^-p$ interaction in baryonic matter relevant to the
in-medium modification of the $\Lambda$(1405) 
resonance also indicate an attractive antikaon potential at $p_K$=0. 

In order to gain some information on the momentum 
dependence of the $K^-$ potential 
we discard the contribution from the unphysical region,
but keep the other terms as fixed by the dispersion 
relation in free space.  In principle, it might 
happen that the $\Lambda$(1405)
is not fully dissolved and might contribute to 
the total amplitude with a negative component.
However, as shown in Fig.~\ref{regen2} and Fig.~\ref{regen2b} 
the amplitude $D$ will be positive above the resonance region 
$p_K{>}$1.5~GeV. Of cause, $D$ may change sign going through a
resonance. Assuming this hypothesis for a moment, one has to
conclude that to saturate both the low energy (kaonic atoms) 
and high energy limits the amplitude $D$
should change its sign at least twice. However, our analysis as well as
the data do not indicate such a strong $K^-$ coupling to resonances
(as compared to the non-resonant contributions) which could 
produce this double oscillation in momentum of 
the real forward scattering amplitude.

\begin{figure}[h]
\vspace{-10mm}
\psfig{file=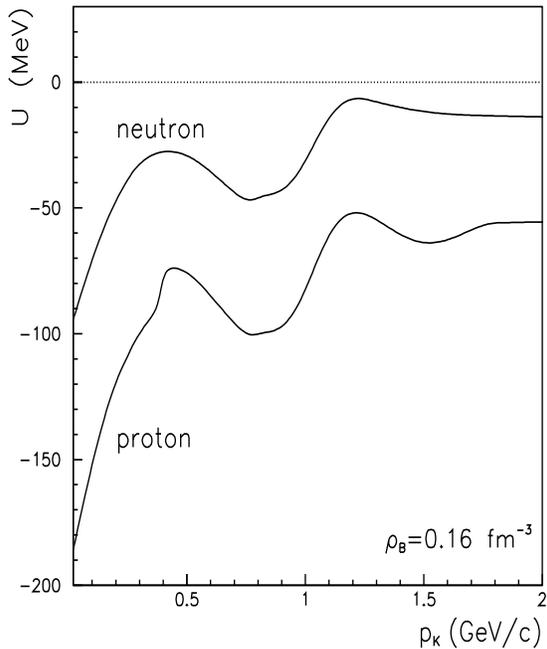,width=8.1cm,height=10cm}
\vspace{-10mm}
\caption{The antikaon potential at $\rho_B$= 0.16~fm$^{-3}$
as a function of the $K^-$ momentum calculated for proton and 
neutron nuclear matter.}
\label{regen8}
\vspace{-5mm}
\end{figure}

Fig.~\ref{regen8} shows the antikaon
potential for the  proton and neutron nuclear matter calculated at
$\rho_B$=0.16~fm$^{-3}$ and averaged over the Fermi 
distribution with Fermi momentum $p_F{\approx}$ 260 GeV/c. 
At $p_K$=0 our result is in line with calculations from 
different dynamical models~\cite{Models}. With increasing 
antikaon momentum the potential increases but remains
attractive. For neutron matter it is smaller than for isospin
symmetric matter as obtained from averaging over isospin.

In summary, within the experimental uncertainties 
our result is in agreement 
with the data on kaonic atoms~\cite{Gal} as well as heavy-ion 
collisions~\cite{IonsT,IonsE}. Furthermore, the FHS Collaboration
\cite{Kiselev} recently reported data on $K^-$ production from $p{+}Be$
collisions where antikaons with $p_K$=1.28~GeV/c were detected 
as a function of the beam energy. The analysis of the data
indicates~\cite{Kiselev} an attractive $K^-$ potential of 
$\simeq$ ${-74}\div$198~MeV for $p_K$=1.28 GeV/c, 
which is also compatible with our
calculation, however, excludes a repulsive potential. 

As proposed in Ref.~\cite{Sibirtsev} the study of the 
antikaon potential in proton-nucleus collisions has a number of 
advantages. First, the $K^-$ production process is well
under control and restricted to a few production channels.
Note, that $\pi\Lambda{\to}K^-X$ and $\pi\Sigma{\to}K^-X$ 
channels, which are important for heavy-ion reactions, do not 
contribute in $p{+}A$ collisions. Second, the measurements can be 
performed in a wide range of momenta $p_K$ relative to the nuclear matter
(which in case of $p{+}A$ interactions is the $K^-$ laboratory
momentum). As indicated the range
$p_K{<}$1.5~GeV/c is very crucial to check if the 
amplitude $D$ may change 
its sign and eventually become negative. These 
advantages are favorable for
the COSY/ANKE experimental project~\cite{ANKE}, which should 
provide further detailed experimental 
information on the in-medium properties of 
the $K^-$-meson at nuclear matter density.

\vspace{0.5cm}
The authors like to acknowledge stimulating discussions with E. L.
Bratkovskaya, C. Fuchs, Y. Kiselev, M. Lutz, U. Mosel 
and H. Str\"oher.


\begin{thebibliography}{9}
\bibitem{Gal}
        E. Friedman, A. Gal and G. Batty, Phys. Lett. B 308 (1993) 6;
        Nucl. Phys. A 579 (1994) 518; E. Friedman, A. Gal,
        J. Mare\u{s} and A. Ciepl\'{y}, nucl-th9804072.
\bibitem{IonsT}
        G.Q. Li, C. M. Ko and X.S. Fang, Phys. Lett. B 329 (1994) 149;
        W. Cassing et al., Nucl. Phys.  A 614 (1997) 415;
        G.Q. Li, C.H. Lee and G.E. Brown, Nucl. Phys. A 625 (1997) 372;
        E. L. Bratkovskaya, W. Cassing and U. Mosel, Nucl. Phys.
        A 622 (1997) 593; Phys. Lett. B 424 (1998) 244; 
        W. Cassing and E.L. Bratkovskaya, Phys. Rep. 308 (1999) 65. 
\bibitem{IonsE}
        A. Schr\"oter et al., Z. Phys. A 350 (1994) 101;
        P. Senger et al., Acta Phys. Pol. B 27 (1996) 2993;
        R. Barth et al., Phys. Rev. Lett. 78 (1997) 4007;
        F. Laue et al., Phys. Rev. Lett. 82 (1999) 1640;
        P. Senger and  H. Stroebele, J. Phys. G 25 (1999) R59. 
\bibitem{Sibirtsev}
        A. Sibirtsev and W. Cassing, Nucl. Phys. A 641 (1998) 476.
\bibitem{PDG}
        Landolt-B\"ornstein, New Series, ed. Schopper, I/12 (1988);
        Particle Data Group, Eur. Phys. J. A 3 (1998) 1.
\bibitem{Theory}
        N.M. Queen, Nucl. Phys. B 1 (1967) 207;
        R. Perrin and W.S. Woolcoock, Nucl. Phys. B 4 (1968) 671;
        P. Bailon et al., Phys. Lett. 50 B (1974) 377;
        O.V. Dumbrais, T.Yu. Dumbrais and N.M. Queen,
        Nucl. Phys. B 26 (1971) 497;
        R. E. Hendrick and B. Lautrup, Phys. Rev. D 11 (1975) 529;
        A. D. Martin, Nucl. Phys. B 179 (1981) 33;
        R. H. Dalitz, Eur. Phys. J. A 3 (1998) 676.
\bibitem{Matrix}
        J.K. Kim, Phys. Rev. Lett. 14 (1965); 19 (1967) 1079;
        B. R. Martin and M Sakitt, Phys. Rev. 183 (1969) 1345;
        A. D. Martin, Nucl. Phys. B 16 (1970) 479.
\bibitem{Data} 
        B. Conforto et al., Nucl. Phys. B 8 (1968) 263;
        R. Armenteros et al., Nucl. Phys. B 21 (1970) 13;
        O.V. Dumbrais, T.Yu. Dumbrais and N.M. Queen,
        Fortschr. Phys. 19 (1971) 491.
\bibitem{Brown}
        G.E. Brown and M. Rho, Nucl. Phys. A 596 (1996) 503;
        T. Waas and W. Weise, Nucl. Phys.  A 625 (1997) 287.
\bibitem{Dynamic}
        V. Koch, Phys. Lett. B 337 (1994) 7;
        T. Waas, M. Rho and W. Weise, Nucl. Phys. A 617 (1997) 449;
        A. Ohnishi, Y. Nara and W. Koch, Phys. Rev. C 56 (1997) 2767.
        M. Lutz, Phys. Lett. B 426 (1998) 12.
\bibitem{Models}
        G.E. Brown and M. Rho, Phys. Rev. Lett. 66 (1991) 2720;
        T. Waas, N. Kaiser and W. Weise, Phys. Lett. B 365 (1996) 12;
        B 379 (1996) 34;
        K. Tsushima, K. Saito, A.W. Thomas and S. Wright,
        Phys. Lett B 429 (1998) 239.
\bibitem{Kiselev}
        Yu. T. Kiselev et al., J. Phys. G 25 (1999) 381.
\bibitem{ANKE}
        T. Kirchner et al., COSY Proposal 21;
        K. Sistemich et al., 'Strangeness in nuclei', Cracow (1992) 359;
        O.W.B. Schult et al.,  Nucl. Phys. A 583 (1995) 629c;
        A. Sibirtsev, H. M\"uller and C. Schneidereit, Z. Phys.
        A 351 (1995) 333;
        M. B\"uscher et al., Acta Phys. Pol. 27 (1996) 3087.
\end{thebibliography}
\end{document}